**Magnetoconductance Oscillations in High-Mobility Suspended Bilayer and Trilayer Graphene**

Wenzhong Bao[1], Zeng Zhao[1], Hang Zhang[1], Gang Liu[1], Philip Kratz[1], Lei Jing[1], Jairo Velasco Jr[1], Dmitry Smirnov[2,] Chun Ning Lau[1*]

[1]Department of Physics, University of California, Riverside, Riverside, CA 92521
[2]National High Magnetic Field Laboratory, Tallahassee, FL 32310

We report pronounced magnetoconductance oscillations observed on suspended bilayer and trilayer graphene devices with mobilities up to 270,000 cm$^2$/Vs. For bilayer devices, we observe conductance minima at all integer filling factors $\nu$ between 0 and -8, as well as a small plateau at $\nu$=1/3. For trilayer devices, we observe features at $\nu$=-1, -2, -3 and -4, and at $\nu$~0.5 that persist to 4.5K at $B$=8T. All of these features persist for all accessible values of $V_g$ and $B$, and could suggest the onset of symmetry breaking of the first few Landau (LL) levels and fractional quantum Hall states.

---

[*] Email: lau@physics.ucr.edu

The fractional quantum Hall effect (QHE), in which the electronic excitations consist of fractionally charge quasiparticles, is an archetypal manifestation of strong electronic interactions in a two-dimensional system. With its anomalous "half-integer" QHE, graphene has emerged as a new platform for physics in low dimensions and with special SU(4) symmetry groups[1, 2]. Spin- and valley-resolved integer[3, 4] and fractional QHE (FQHE) [5, 6] have been observed in single layer graphene (SLG). Very recently, bilayer graphene (BLG) and trilayer graphene (TLG) have also attracted significant attention; many novel phenomena are predicted, such as tunable band gap[7], tunable excitons[8] with possibility of Bose-Einstein condensation[9], and unusual flavor symmetry[10]. Integer QHE that are orbital-, spin- and valley-resolved have been observed in suspended[11] and substrate-supported[12] BLG devices. However, even though the eight-fold degeneracies at zero energy are completely lifted at relatively low magnetic field $B$=3T, no FQHE was observed up to 12T[11]. For TLG devices, which have much lower mobility, even four-fold degenerate QHE has not been observed.

Here we present low temperature transport measurements of magnetoconductance (MC) on suspended BLG and TLG devices with mobilities up to 270,000 cm$^2$/Vs. Shubnikov de Haas (SdH) oscillations appear at magnetic fields as low as 0.2 T. The devices' MC exhibit pronounced dips at integer values of filling factor $\nu$ that are constant with $n/B$, where $n$ is the induced charge density and $B$ is the magnetic field. For BLG devices, such dips for $0 \geq \nu \geq -8$ are clearly resolved at $B$<3T, resolved in the same order as QH plateaus in previous reports[11, 12]. We identify these dips as signatures of degeneracy-lifted QH states, and attribute the lack of conductance quantization to the two-terminal geometry and the presence of strains and/or ripples[13] that may induce local gauge fields[14-16]. At high magnetic fields up to 31T, we observe a plateau-like feature at $\nu$=1/3 that scales appropriately with $n$ and $B$ and disappears at temperature $T$~2-5K, which could suggest the onset of a FQH state in bilayer devices.

For TLG devices, we observe similar MC dips at integer values of $\nu$, for $0 \geq \nu \geq -4$, and a feature at filling factor $\nu$=0.5±0.07 that persists up to 5K, which may correspond to the $\nu$=1/2 or 2/5 FQH state. The different $T$ dependence of these features at fractional filling factors in bilayer and trilayer devices may reflect the different energy gaps and electronic interaction strengths of these atomic membranes.

The graphene devices, with typical areas ~10-100 μm$^2$ (Fig. 1 inset), are fabricated by exfoliating graphite over pre-defined trenches on doped Si/SiO$_2$ substrates, and depositing electrodes via shadow mask evaporation[17]. They are measured at low temperatures using standard lock-in techniques. The blue curves of Fig. 1 display the two terminal conductance $G$ vs. gate voltage $V_g$, for as-fabricated BLG (left panel) and TLG (right panel) devices at 4.2 K; their Drude mobilities $\mu_D = \sigma/ne$ are typically ~10,000-30,000 cm$^2$/Vs, where $\sigma$ is the two terminal device conductivity and $e$ is the electron charge. After current annealing[18] at ~ 0.1 – 0.2 mA/μm/layer, the $G(V_g)$ characteristics display much sharper Dirac points that are closer to zero (red curves, Fig. 1). For a typical post-annealed BLG device, $\mu_D$ ranges from 100,000 to 274,000 cm$^2$/Vs at $n$~10$^{10}$ cm$^2$, while their field effect mobility $\mu_{FE} = \frac{1}{e}\frac{d\sigma}{dn}$ ranges from 28,000 to 200,000. For TLG, $\mu_{FE}$ and $\mu_D$ are ~50,000 and 200,000 cm$^2$/Vs, respectively. These values are exceedingly high, especially considering that the mobility of a BLG device is typically an order of magnitude lower than that of SLG, and that of a TLG device is less than 1,000. Thus, both the mobility values and the device areas of our devices are significantly larger than those previously reported[11, 12].

In magnetic fields, the LL energies of BLG are $E_N^{BL} = \pm \frac{\hbar eB}{m^*}\sqrt{N(N-1)}$ [19, 20], where $m^* \sim 0.04 m_e$ is the effective mass of its charge carriers and $m_e$ is electron's rest mass. When the Fermi level is between the LLs, the device's Hall conductance is expected to be quantized at $\sigma_{xy}^{BL} = 4N\frac{e^2}{h}$, where $N=$... -3, -2, -1, 1, 2, 3... is an integer denoting the LL index and $h$ Planck's constant. The $N=1$ and $N=0$ LLs are doubly degenerate, resulting in an eight-fold degeneracy at zero energy.

We now examine the conductance of a bilayer device BL1 in finite $B$ at $T=300$mK (similar data were observed on 2 other samples). As shown by Fig. 2a, which plots $G$ vs. $1/B$ at 10 different gate voltages, $G$ displays pronounced SdH oscillations, which are discernible at $B$ as low as 0.2 T. The exceedingly high mobility of the devices, together with the low field at which SdH oscillations become visible, underscore the high quality of our devices. Yet, the device conductance is not properly quantized, even at the highest attainable magnetic field. This absence of quantization is not fully understood, but could be attributed to 3 factors. Firstly, due to the device's two-terminal geometry, $G$ comprises of both longitudinal and transverse contributions[21], thus displaying non-monotonous dependence on $n$, and non-quantized conductance for sufficiently broadened LLs. Another possible reason is the presence of strain and/or ripples in our devices[13] that have suspended portions up to 5 µm long and rests on the rigid banks of the trenches. Their deflection under applied $V_g$, which scales with the 4/3 power of the length, could produce significant strain close to their rigid boundaries, which in turn result in gauge fields that partially destroy the conductance quantization[14, 22]. A third possible factor is the small substrate-supported area of the device (typically <10% of the total device area), which presumably has lower mobility and may not exhibit QHE at low $B$, thus destroying the overall conductance quantization.

Despite the lack of conductance quantization, it is possible to extract information on QH states from the data. For this device that is short and wide, filling factors $v$ of the conductance minima can be used to identify QH features[21]. To this end, we note that $B_F = nh/4e = (\alpha V_g)h/4e$, where $1/B_F$ is the period of the SdH oscillations, and $\alpha = n/V_g$ is the coupling efficiency of the back gate (here $V_g$ is measured from the Dirac point). Plotting the measured values of $B_F$ vs. $V_g$ indeed yields a straight line, with a best-fitted slope of $a=0.26$ T/V (Fig. 2b). This indicates $\alpha = a(4e/h) \approx 2.5 \times 10^{10}$ cm$^{-2}$ V$^{-1}$, in agreement with that independently estimated from the device geometry. We can thus *unequivocally* determine the filling factor corresponding to any given data point,

$$v = nh/Be = 4a(V_g/B) \approx 1.05(V_g/B) \quad (1)$$

To examine the data more closely, we plot $G(V_g)$ at different values of $B$ (Fig. 2c). The conductance exhibits pronounced oscillations, with the minima occurring at $V_g$ that correspond to integer $v$ that is calculated using Eq. (1). For instance, clear conductance minima for $-4 \leq v \leq 0$ are visible at $B=2$T, and resolved successively in the order $v=-4, 0, -2, -3, -1$. This is reminiscent of the data reported in ref. [11, 12], in which the $v=0$ QH plateau appears at the lowest field, followed by the plateaus $v=2, 3$ and $1$. Hence, the observation of conductance minima at integer filling factors, and their resolution in the same order as in previous reports, suggest that these minima arise from integer QH effect in BLG, with the orbital, spin and valley degeneracies lifted.

Despite these suggestive observations, we caution that QH features cannot be inferred from a *single* $G(V_g)$ curve, as features coincide with apparent integer $\nu$ may arise from inadvertent formation of *pn* junctions[23-26], where $\nu$ of the differently doped regions are not known, or from localization-induced fluctuations[26]. However, as we can unambiguously determine $\nu$ for each feature, such ambiguity can be removed by plotting $G$ *vs.* both $V_g$ and $B$, since a QH plateau with a given $\nu$ will have a slope $\nu e/h\alpha$ in the $V_g$-$B$ plane, regardless of its actual conductance value[26]. In the absence of LL splitting, we expect to observe plateaus only with slopes of ±4, ±8, etc for BLG[27].

To verify this, we plot the evolution of $G(V_g, B)$ (upper panel) and $dG/dV_g$ (lower panel) in Fig. 2d. The bands of colors that radiate from $V_g$~ 0.3V, which is inferred to be the Dirac point, mark the onset of SdH oscillations. The MC can be seen more clearly by differentiating $G$ with respect to $V_g$, where the blue (red) regions indicate negative (positive) values of $dG/dV_g$; the local conductance minima appear as white regions in the $V_g$-$B$ plane, as outlined by the dotted lines. Strikingly, from their slopes in the $V_g$-$B$ plane and Eq. (1), the filling factors of these minima are identified to span all integers between 0 and -8 when the device is hole-doped (albeit the $\nu$=-5 and -6 minima are just barely distinguishable); for the electron doped regime, because of the limited $V_g$ range, only minima with $\nu$=2 and 4 are identified. Such persistence of these conductance dips at integer values of $\nu$, which are observed for *all* accessible values of $V_g$ and $B$, provides very strong evidence that they indeed arise from orbital-, spin- and valley-resolved QH states.

We note that this is the first report of possible symmetry breaking for the $N$=2 LL, which is expected to exhibit interaction effects. For instance, we observe that the $\nu$=7 state is resolved before the $\nu$=5, 6 states, suggesting a larger energy gap for the former. This is quite surprising, since the even integer states are expected to be resolved first. Further investigation would be necessary to provide further insight into these symmetry-broken higher LL states in BLG devices.

We now focus on the BLG device behavior in higher fields 4<$B$<31T. To avoid collapsing of the atomic membranes, we restrict the applied $|V_g|$ to <10V; thus, for $B$>10T only QH state with $|\nu|$ <1 are experimentally accessible. In the $G(V_g, B)$ plot (Fig. 3a), a white/pink feature with a shallow slope is discernible. Its slope in the $V_g$-$B$ plane is $V_g/B$~0.32, yielding $\nu$ ≈0.33. This feature can be seen more clearly by taking discrete line traces at different $B$ values (Fig. 3b) – it appears as a broadened peak for $B$<15T, but develops into a small plateau with increasing $B$. Fig. 3c replots these traces as $G(\nu)$, where $\nu$ is calculated using Eq. (1) with the small offset in Dirac point taken into account. As expected, for B>15T, the traces nearly collapse into one, with the small plateau located at $\nu$=0.33.

Taken together, our data are suggestive of signatures of the fractional $\nu$=1/3 QHE state in BLG. In previous works on SLG, the $\nu$=1/3 fractional state is surprisingly robust and persists up to 20K at $B$=12T, with a large, Coulomb interaction-induced energy gap $\Delta_{1/3}^{SL}$~ 10 K·$\sqrt{B}$[6]. In contrast, there is little theoretical effort on fractional QHE in BLG[28]. Taking the features in our data as an evidence for the 1/3 FQH state in bilayer, we can obtain an order-of-magnitude estimate for $\Delta_{1/3}^{BL}$ by measuring $G(V_g)$ at several different temperatures $T$ (Fig. 3d). At $B$=20T, the small 1/3 plateau persists at $T$=1.3K, but disappears completely at $T$=5.5K, yielding an estimated $\Delta_{1/3}^{BL}$~ 0.4 K·$\sqrt{B}$, which is much smaller than that of SLG. The increase in the overall conductance with $T$ also suggests the presence of significant thermally activated conduction through the bulk of the device.

Finally, we turn our attention to TLG, which are assumed to be Bernal-stacked. Tight binding calculations predict that the energy spectrum of a Bernal-stacked TLG is a superposition of those of single layer and bilayer graphene[29, 30]. Similar to BLG, we expect that its spin, valley and layer degeneracies can be broken by $B$ or electronic interactions. However, no QHE of any type has been observed in TLG to date.

Our experimental measurements on a trilayer device with $\mu_{FE} \sim \mu_D \sim 50,000$ cm$^2$/Vs at 260mK reveal pronounced MC oscillations. Using the slopes of the conductance features in the $V_g$-$B$ plane, we identify QH features at $\nu$=0, 11±1, -4, -2, -3 and -1, which are resolved in the order listed. We note that, albeit without proper quantization, the conductances are within ~30% of the expected values. Fig. 4a displays $G(V_g, B)$ for 2≤$B$≤8T for such a device TL1. At Dirac point $V_g$~0.6V, $G$ decreases steadily with $B$, and becomes limited by instrument noise floor (<5nS) for $B$>6T, suggesting the presence of an insulating state at $\nu$=0. Line traces $G(V_g)$ at different $B$ values are shown in Fig. 4b,d. By plotting the same data as $G(\nu)$, these traces collapse into a single curve (Fig. 4c,e), with plateaus or shoulders at integer $\nu$. In particular, Fig. 4e exhibit two identifiable features: "A" that appears at $\nu$=-1.0±0.03, and "B" at $\nu$=0.50±0.07. Both features are relatively robust in temperature and persist up to 4.5K (Fig. 4f).

Feature "B" is particularly intriguing, since it may correspond to the $\nu$=1/2 or 2/5 state. A similar feature in SLG has been observed[5], yet its origin is still under debate, since the $\nu$=1/2 feature in $R_{xx}$ in traditional GaAs devices arises from a Fermi liquid state, not FQHE[31]. However, we note that a $\nu$=1/2 FQH state is observed in *bilayer* GaAs devices[32, 33]; thus, though not conclusive, feature "B" may in fact indicate a FQH state in TLG with a relatively large energy gap. The absence of the $\nu$=1/3 state may be attributed to the presence of the $\nu$=0 insulating state, which, if sufficiently wide, is shown to mask signatures of FQH states in suspended single layer devices[6].

We thank M. Bockrath, N. Kalugin, H. Fertig and E. McCann for discussions and T. Murphy for assistance at NHMFL. We acknowledge the support of NSF CAREER DMR/0748910, NSF/ECCS 0926056, ONR N00014-09-1-0724, and FENA Focus Center. D.S. acknowledges the support by NHMFL UCGP #5068. Part of this work was performed at NHMFL that is supported by NSF/DMR-0654118, the State of Florida, and DOE.

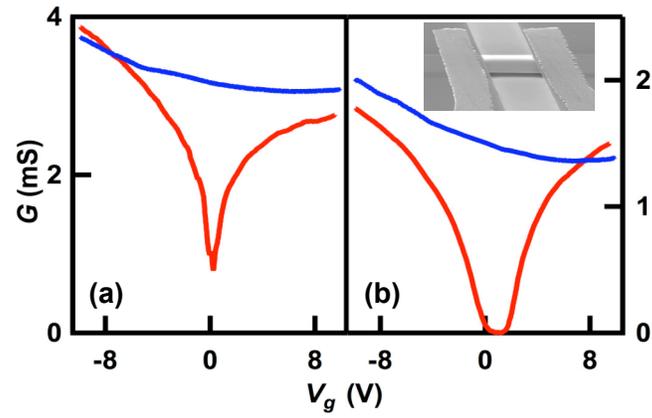

Fig. 1. $G(V_g)$ for **(a)** BLG and **(b)** TLG devices at $T$=4.2K. Blue and red curves are taken before and after current annealing, respectively. Inset: SEM image of a suspended graphene device.

Fig. 2. Data from a bilayer device BL1 at 300mK. **(a).** $G$ vs. $1/B$ at $V_g$=3, 4, 5, 6,7, 7.5, 8, 8.5, 9 and 9.8V (bottom to top). The traces are offset for clarity. **(b).** $B_F(V_g)$ and a linear fit to the data points. **(c).** $G(V_g)$ at B=3, 2 and 1.5T (bottom to top). The traces are offset for clarity. The numbers indicate the $|\nu|$ values that correspond to the local conductance minima. **(d).** $G$ in $e^2/h$ (upper) and $dG/dV_g$ (lower panel) vs $B$ and $V_g$. The dotted lines correspond to features with integer $\nu$ between $\nu$=0 and -8.

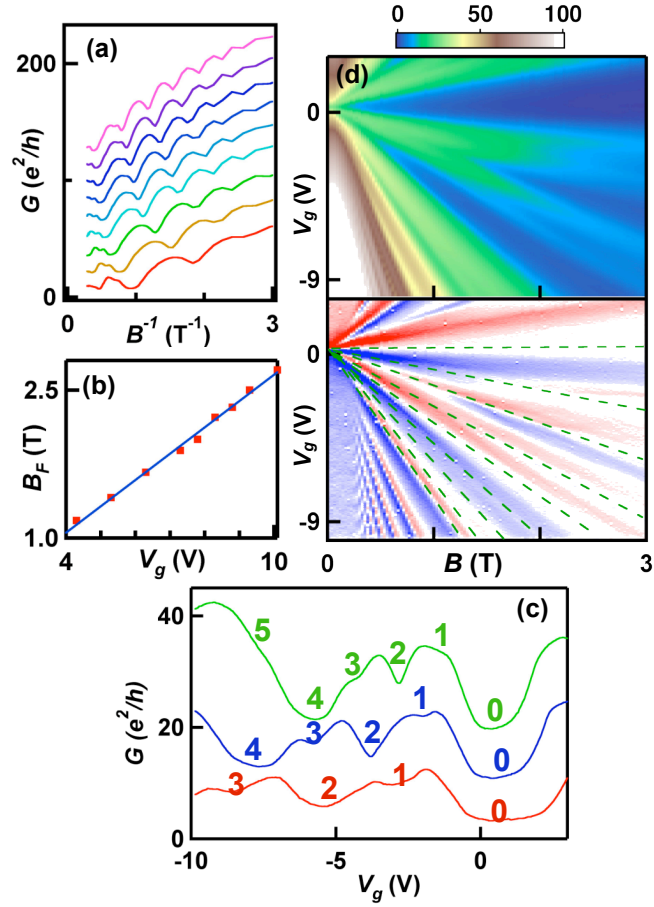

Fig. 3. High field data for BL1. **(a)**. $G(V_g, B)$ at 300 mK. **(b-c)**. Line traces from (a) at $B$=15T, 17.5T, 20T, 23T and 28.5T (right to left in (b)), plotting against $V_g$ and $\nu$, respectively. **(d)**. $G(V_g)$ at $B$=20T and different $T$. The traces are *not* offset.

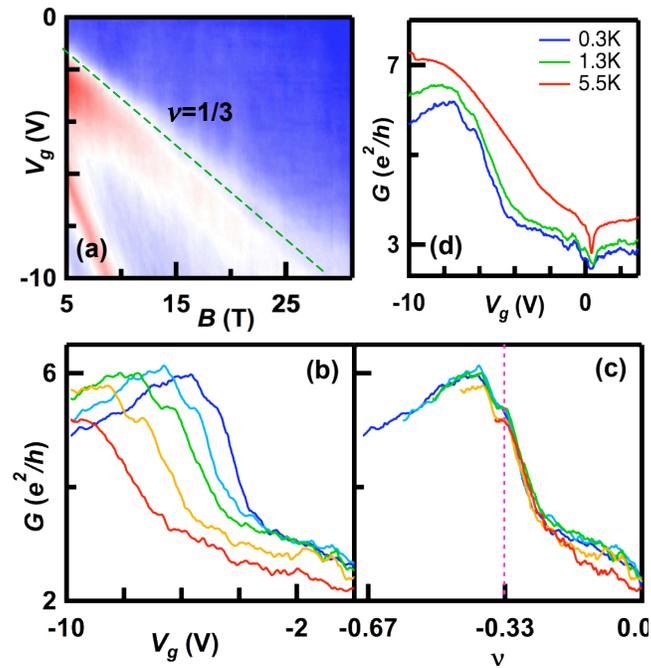

Fig. 4. Data from a trilayer device TL1. **(a).** $G(V_g, B)$ at 260mK. **(b-c).** $G(V_g)$ and $G(\nu)$ at $B=2.2$, 2.5, 3.0, 3.5, 3.8T. **(d-e).** $G(V_g)$ and $G(\nu)$ at $B=4, 5, 6, 7$ and 8T. **(f).** $G(V_g)$ at $B=8T$ and $T=4.5$, 2.7, 1.9, 1.6, 1.3, 1.0, 0.7, 0.4 and 0.26 K(top to bottom). The traces are offset for clarity.

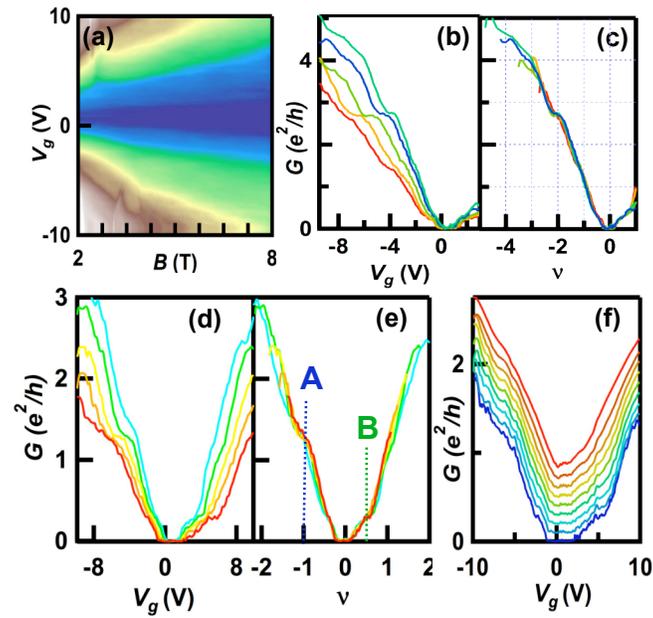